\documentclass[aps,prc,twocolumn,showpacs,preprintnumbers,amsmath,amssymb,
epsf,superscriptaddress,groupedaddress,nofootinbib,tightenlines]{revtex4}

\usepackage{graphicx}
\usepackage{amsmath}
\usepackage{bm}

\begin{document}

\title{Model Independent Naturalness Bounds on Magnetic Moments of Majorana Neutrinos}

\author{Mikhail Gorchtein}

\author{Nicole F. Bell}

\author{Michael J. Ramsey-Musolf}

\author{Petr Vogel}

\author{Peng Wang}
  
\affiliation{California Institute of Technology, Pasadena, CA 91125, USA}

\begin{abstract}
We analyze the implications of neutrino masses for the magnitude of
neutrino magnetic moments.  By considering electroweak radiative
corrections to the neutrino mass, we derive model-independent
naturalness upper bounds on neutrino magnetic moments,
generated by physics above the electroweak scale. For 
Majorana neutrinos, these bounds are weaker than
present experimental limits if $\mu_\nu$ if generated by new physics
at $\sim$ 1 TeV, and surpass current experimental sensitivity only for
new physics scales $>$ 10 -- 100 TeV. The discovery of a neutrino
magnetic moment near present limits would thus signify that neutrinos
are Majorana particles.
\end{abstract}

\maketitle


In the Standard Model (minimally extended to include non-zero neutrino
mass) the neutrino magnetic moment is given by
$\mu_\nu\approx 3\times 10^{-19}\left(\frac{m_\nu}{1{\rm eV}}\right)\mu_B$~\cite{Marciano:1977wx}. 
An experimental observation of a larger magnetic moment 
would be an unequivocal indication of physics
beyond the minimally extended Standard Model.  Current laboratory
limits, $\mu_\nu < 1.5 \times 10^{-10} \mu_B$~\cite{Beacom} and
$\mu_\nu < 0.7 \times 10^{-10} \mu_B$~\cite{reactor}, are obtained from
solar and reactor experiments, respectively.  Slightly stronger bounds
are obtained from astrophysics, $\mu_\nu < 3 \times 10^{-12}$~\cite{Raffelt}.  
In the case of Dirac neutrinos, the neutrino mass constrains the magnetic 
moments to be~\cite{dirac} 
$\mu_\nu \leq 10^{-15} \mu_B$, which is several orders of
magnitude more stringent than current experimental constraints.

Following Refs.~\cite{dirac,Davidson} we assume that the magnetic
moment is generated by unspecified physics beyond the SM at an energy scale
$\Lambda$ above the electroweak scale.  We shall work exclusively with 
dimension $D>4$
operators that involve only SM fields, obtained by integrating out the
physics above the scale $\Lambda$, and respect the
$SU(2)_L\times U(1)_Y$ symmetry of the SM, and contains only SM fields
charged under these gauge groups. 
The lowest order contributions to the neutrino (Majorana) mass arise
from the usual five and seven dimensional operators containing Higgs and
left-handed lepton doublet fields, $L$ and $H$, respectively, 
${\bf \left[O_M^{5D}\right]_{\alpha\beta}=\left(\overline{L^c_\alpha}\epsilon H\right)\left(H^T\epsilon L_\beta\right)}$ the five dimensional, and 
${\bf\left[O_M^{7D}\right]_{\alpha\beta}=\left(\overline{L^c_\alpha}\epsilon H\right)\left(H^T\epsilon L_\beta\right) \left(H^\dagger H \right)}$ the seven 
dimensional one.
We use the notation $\epsilon = - i \tau_2$, $\overline{L^c}=L^TC$, $C$ denotes
charge conjugation, and $\alpha$, $\beta$ are flavor indices.  
The neutrino magnetic moment operator is
generated by dimension 7 operators involving the $SU(2)_L$ and
$U(1)_Y$ gauge fields, $W_\mu^a$ and $B_\mu$, respectively, 
${\bf\left[O_B\right]_{\alpha\beta}= g'\left(\overline{L^c}_\alpha\epsilon H\right)\sigma^{\mu\nu}\left(H^T\epsilon L_\beta\right)B_{\mu\nu}}$, and 
${\bf\left[O_W\right]_{\alpha\beta}= g\left(\overline{L^c_\alpha}\epsilon H\right)\sigma^{\mu\nu}\left(H^T\epsilon \tau^a L_\beta\right)W_{\mu\nu}^a}$. 
In the above definitions, 
$B_{\mu\nu} = \partial_\mu B_\nu - \partial_\nu B_\mu$ and
$W_{\mu\nu}^a = \partial_\mu W_\nu^a - \partial_\nu W_\mu^a - g
\epsilon_{abc}W_\mu^b W_\nu^c$ are the U(1)$_Y$ and SU(2)$_L$ are the field
strength tensors, respectively, and $g'$ and $g$ are the corresponding
couplings. The three 7D operators defined above do not 
form a basis under renormalization.  The
full basis of 7D operators may be found in \cite{majorana}.

The operator $O_W$ is asymmetric in flavor, and it is useful to
express it in terms of operators with explicit flavor symmetry,
$\left[ O_W^\pm \right]_{\alpha\beta}=\frac{1}{2} (\left[O_W\right]_{\alpha\beta} \pm \left[O_W\right]_{\beta\alpha} )$.
Operators $O_M^{5D}$, $O_M^{7D}$ and $O_W^+$ are flavor symmetric, 
while $O_B$ and $O_W^-$ are antisymmetric. 
Our effective Lagrangian is therefore
\begin{eqnarray}
{\cal L} &=& \frac{C_M^{5D}}{\Lambda} O_M^{5D} 
+ \frac{C_M^{7D}}{\Lambda^3} O_M^{7D}+  \frac{C_{B}}{\Lambda^3} O_{B} 
+\frac{ C_{W}^+}{\Lambda^3} O_{W}^+ \nonumber\\
&+&  \frac{C_{W}^-}{\Lambda^3} O_{W}^-
+\dots ,
\end{eqnarray}
where the \lq\lq $+\cdots$" denote other terms that are not relevant to the 
present analysis.
After spontaneous symmetry breaking, the above operators reduce to the 
flavor symmetric Majorana mass term 
${\bf \frac{1}{2}\left[m_\nu\right]_{\alpha\beta}\overline{\nu^c}_\alpha \nu_\beta}$,
and flavor antisymmetric Majorana magnetic moment interaction 
${\bf\frac{1}{2} \left[\mu_\nu\right]_{\alpha\beta}\, \overline{\nu^c}_\alpha \sigma^{\mu\nu}\nu_\beta F_{\mu\nu}}$, where $F_{\mu\nu}$ is the electromagnetic 
field strength tensor. At the EW scale, we have
\begin{eqnarray}
\frac{\left[\mu_\nu\right]_{\alpha\beta}}{\mu_B}&=&\frac{2m_e v^2}{\Lambda^3} 
\left(\left[C_B(M_W)\right]_{\alpha\beta} 
+ \left[C_W^-(M_W)\right]_{\alpha\beta}\right),\nonumber\\
&&\label{munu}\\
\frac{1}{2}\left[ m_\nu \right]_{\alpha\beta} 
&=&\frac{v^2}{2 \Lambda} \left[C_M^{5D}(M_W)\right] 
+ \frac{v^4}{4 \Lambda^3} \left[C_M^{7D}(M_W)\right],
\label{mnu}
\end{eqnarray}
\indent
where the Higgs VEV is $< H^T>=(0,v/\sqrt{2})$.

Below, we calculate radiative
corrections to the neutrino mass operators ($O_M^{5D}$ and $O_M^{7D}$)
generated by the magnetic moment operators $O_W^-$ and $O_B$ and put 
constraints on the size of the magnetic moment in terms of the neutrino mass, 
using Eqs.(\ref{munu}) and (\ref{mnu}).  

\begin{itemize}
\item {\bf 5D mass term - 1 loop}
\end{itemize}
One-loop radiative corrections only yield a contribution to $O_M^{5D}$ 
associated with the flavor symmetric operator  $O_W^+$. No model independent
bound can be obtained from the one-loop radiative corrections from magnetic 
momentto the 5D mass operator. 

\begin{figure}[ht]
\includegraphics[width = 1in]{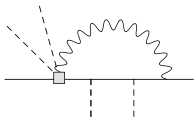}
\caption{Contribution of $O_W^-$ to the 7D neutrino mass operator.}
\label{fig:7D}
\end{figure}

\begin{itemize}
\item {\bf 7D mass term - one loop}
\end{itemize}
As the operator $O_W^-$ is flavor antisymmetric, it must be
multiplied by another flavor antisymmetric contribution in order to
produce a flavor symmetric mass term.  This can be accomplished
through insertion of Yukawa couplings in the diagram shown in
Fig.~\ref{fig:7D}~\cite{Davidson}.  
This diagram provides a
logarithmically divergent contribution to the 7D mass term, given by
\begin{equation}
\label{eq:owminusone}
\left[ C_M^{7D}(M_W)  \right]_{\alpha\beta}
\simeq \frac{3 g^2}{16 \pi^2}
\frac{m_\alpha^2 - m_\beta^2}{v^2} 
\ln \frac{\Lambda^2}{M_W^2}
\left[ C_W^-(\Lambda) \right]_{\alpha\beta}
\end{equation}
where $m_\alpha$ are the charged lepton masses, and the exact
coefficient has been computed using dimensional regularization and renormalized 
with modified minimal subtraction. 
Using this result, as well as Eqs. (\ref{munu}) and (\ref{mnu}), leads to 
bound (i) in Table~\ref{summary}.

\begin{figure}[ht]
\includegraphics[width = 1in]{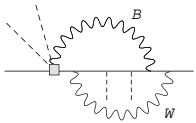}
\caption{Representative contribution of $O_B$ to the 7D neutrino
mass operator at two loop order.}
\label{fig:B2M_2loop}
\end{figure}

If we insert $O_B$ in the diagram in Fig. \ref{fig:7D}, the
contribution vanishes. A non-zero contribution to $O_M^{7D}$ from
$O_B$ can arise at two-loop level, for instance, from a virtual $W$ boson 
loop as in Fig. \ref{fig:B2M_2loop}~\cite{Davidson}.
This mechanism gives the leading contribution of the operator $O_B$ to
the 7D mass term.   The $O_B$ and $O_W$ contributions to the 7D mass term 
are thus related by $\frac{(\delta m_\nu)^B}{(\delta m_\nu)^W}\approx\frac{\alpha}{4\pi} \frac{1}{\cos^2\theta_W}$, where $\theta_W$ is the weak mixing angle. 
The corresponding limit is shown as bound (iii) in Table~\ref{summary}.

\begin{figure}[h]
\includegraphics[width = 1in]{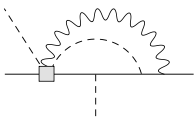}
\caption{Representative contribution of $O_W^-$ and $O_B$ to the 5D neutrino
mass operator.}
\label{fig:2loop}
\end{figure}

\begin{itemize}
\item {\bf 5D mass term - two loop}
\end{itemize}
The neutrino magnetic moment operator $O_W^-$ will also
contribute to the 5D mass operator {\em via} two-loop matching of the 
effective theory onto the full theory at at the scale $\Lambda$, as shown in 
Fig.~\ref{fig:2loop}.
This diagram contributes to
the 5D mass operator, and we provide a
na\"ive dimensional analysis~\cite{Burgess:1992gx} (NDA) estimate for the
$O_B$ and $O_W$ insertions:
\begin{eqnarray}
\left[C_M^{5D}(\Lambda)  \right]_{\alpha\beta}
&\simeq &
\frac{g^2}{(16 \pi^2)^2}
\frac{m_\alpha^2 - m_\beta^2}{v^2} 
\left[  C_W^- (\Lambda)\right]_{\alpha\beta}.
\label{5d1}
\end{eqnarray}
\indent
Using Eqs.(\ref{munu}) and (\ref{mnu}), we obtain the bound
(ii) in Table~\ref{summary}. 
The $O_B$ contribution to the 5D mass term is the same as that for $O_W$,
except for a factor of $(g'/g)^2 = \tan^2 \theta_W$, and it corresponds
to the bound (iv) in Table~\ref{summary}.
In doing so, we have neglected the running of the 
operator coefficients from the scale $\Lambda$ to $M_W$ since the effects are 
higher order in the gauge couplings and have a negligible numerical impact on 
our analysis. 
\begin{table}[htbp]
   \centering
   \begin{tabular}{c|c|c}
\hline\hline
i) 1-loop, 7D & 
$\mu^W_{\alpha\beta}$ & $ \leq 1 \times 10^{-10}\mu_B
\left(\frac{ \left[m_\nu\right]_{\alpha\beta}}{1~{\rm eV}}\right)
\ln^{-1}\frac{\Lambda^2}{M_W^2} R_{\alpha\beta}$ \\
ii) 2-loop, 5D & $\mu^W_{\alpha\beta}$ & $ \leq 1 \times 10^{-9}\mu_B
\left(\frac{ \left[m_\nu\right]_{\alpha\beta}}{1~{\rm eV}}\right)
\left(\frac{1~{\rm TeV}}{\Lambda}\right)^2 
R_{\alpha\beta}$ \\  
\hline
iii) 2-loop, 7D & $\mu^B_{\alpha\beta}$ & $ \leq 1 \times 10^{-7}\mu_B
\left(\frac{ \left[m_\nu\right]_{\alpha\beta}}{1~{\rm eV}}\right)
\ln^{-1}\frac{\Lambda^2}{M_W^2}
R_{\alpha\beta}$ \\
iv) 2-loop, 5D &
$\mu^B_{\alpha\beta}$  &  $\leq 4 \times 10^{-9} \mu_B
\left(\frac{ \left[m_\nu\right]_{\alpha\beta}}{1~{\rm eV}}\right)
\left(\frac{1~{\rm TeV}}{\Lambda}\right)^2 
R_{\alpha\beta}$ \\
\hline\hline
   \end{tabular}
   \caption{Summary of constraints on the magnitude of the magnetic
moment of Majorana neutrinos.  The upper two lines correspond to a magnetic moment generated
by the $O_W$ operator, while the lower two lines correspond to the
$O_B^-$ operator. $R_{\alpha\beta} = \frac{m_\tau^2}{|m_\alpha^2 - m_\beta^2|}$,
with $m_\alpha$ being the masses of charged lepton masses. Numerically, 
$R_{\tau e} \simeq R_{\tau \mu} \simeq 1$ and $R_{\mu e} \simeq 283$.}
   \label{summary}
\end{table}
Our conclusions can be summarized according to the scale of the new
physics, $\Lambda$:
{\flushleft
I) $\Lambda \leq 10$ TeV 
\begin{itemize}
\item 
No conflict with experimental limits.
\item
Both $O_W$ and $O_B$ contributions to $\mu_\nu$ are possible, though
$O_W$ contributions are more tightly constrained.
\end{itemize}
II) $\Lambda \geq 10$ TeV
\begin{itemize}
\item 
$\mu_{\tau\mu}$, $\mu_{\tau e}$ bounds stronger than experimental limits
\item 
$\mu_{\mu e}$ bound weaker than experimental limits 
\item
Same limit irrespective of whether $\mu_\nu$ generated by $O_W$ and $O_B$
\end{itemize}
III) $\Lambda \geq 100$ TeV
\begin{itemize}
\item
The $\mu_{\alpha\beta}$ bound becomes stronger than current experimental constraints, for all $\alpha,\beta$.
\end{itemize}
}
We emphasize that the bound (iv) in Table \ref{summary} is the most general 
bound on the magnetic moment.

We thank Vincenzo Cirigliano, Concha Gonzalez-Garcia, and Mark Wise
for illuminating conversations.  This work was supported in part under
U.S. DOE contracts DE-FG02-05ER41361 and DE-FG03-92ER40701.


\begin{thebibliography}{9}

\bibitem{Marciano:1977wx}
W.~J.~Marciano and A.~I.~Sanda,
Phys.\ Lett.\ B {\bf 67}, 303 (1977);
%
B.~W.~Lee and R.~E.~Shrock,
Phys.\ Rev.\ D {\bf 16}, 1444 (1977);
%
K.~Fujikawa and R.~Shrock,
Phys.\ Rev.\ Lett.\  {\bf 45}, 963 (1980).

\bibitem{Beacom} 
 J.~F.~Beacom and P.~Vogel,
Phys.\ Rev.\ Lett.\  {\bf 83}, 5222 (1999);
D.~W.~Liu {\it et al.},
Phys.\ Rev.\ Lett.\  {\bf 93}, 021802 (2004).


\bibitem{reactor} 
H.~T.~Wong  [TEXONO Collaboration],
  hep-ex/0605006.
B.~Xin {\it et al.}  [TEXONO Collaboration],
  Phys.\ Rev.\ D {\bf 72}, 012006 (2005);
Z.~Daraktchieva {\it et al.}  [MUNU Collaboration],
  Phys.\ Lett.\ B {\bf 615}, 153 (2005).


\bibitem{Raffelt} G.G. Raffelt, Phys. Rep. {\bf 320}, 319 (1999).

\bibitem{dirac}
  N.~F.~Bell, V.~Cirigliano, M.~J.~Ramsey-Musolf, P.~Vogel and M.~B.~Wise,
  Phys.\ Rev.\ Lett.\  {\bf 95}, 151802 (2005).

\bibitem{Davidson}
  S.~Davidson, M.~Gorbahn and A.~Santamaria,
  Phys.\ Lett.\ B {\bf 626}, 151 (2005).

\bibitem{majorana} N.F. Bell, M. Gorchtein, M.J. Ramsey-Musolf, P. Vogel, P. Wang, Phys. Lett. B {it in press}, arXiv:[hep-ph/0606248]

\bibitem{Burgess:1992gx}
  C.~P.~Burgess and D.~London,  
  Phys.\ Rev.\ D {\bf 48}, 4337 (1993).

\bibitem{WMAP3yr}
  D.~N.~Spergel {\it et al.},
  astro-ph/0603449.

\end{thebibliography}
\end{document}